\documentclass[12pt,reqno]{article}
\usepackage{amsmath, amsthm, amssymb,epsfig,alltt}
\usepackage{mathtools}

\def\a{\alpha}

\def\l{\lambda}

\def\p{\partial}

\def\t{\tau}

\def\s{\sigma}
\def\g{\gamma}

\def\half{\frac{1}{2}}

\def\barz{{\bar z}}

\def\sp{\sigma^\prime}
\def\nn{\nonumber}

\def\2pap{2\pi\alpha^\prime}

\def\beq{\begin{eqnarray}}
 \def\eeq{\end{eqnarray}}
 \def\4pap{4\pi\a^\prime}
 
 \def\sp{{\s^\prime}}
 
 \def\tp{{\t^\prime}}
 \def\zp{{z^\prime}}

 \def\barPsi{{\bar \Psi}}
 
 \def\barz{{\bar z}}
 \def\barzp{{\bar z}^\prime}
 
 \def\barxi{{\bar \xi}}

 \def\calO{{\cal O}}
 \def\boleta{\boldsymbol \eta}
 \def\bolPsi{\boldsymbol \Psi}

 \def\bolR{\boldsymbol R}
 \def\bolM{\boldsymbol M}
 \def\bolD{\boldsymbol D}
 
\begin{document}


\title{Resonant Multilead Point-Contact Tunneling: \\
Boundary State Formulation }

\author{Taejin Lee \\~~\\
Department of Physics, Kangwon National University, \\
Chuncheon 24341 Korea \\
email: taejin@kangwon.ac.kr}

\maketitle

\centerline{\bf Astract}
We study a model of resonant multilead point-contact tunneling by using the boundary state formulation.
At a critical point the model is described by multi-flavor chiral fermions on an infinite line with a point 
contact interaction at the origin. By applying the folding procedure, previously developed for the model 
of resonant point-contact tunneling of a single lead, we map the model onto a non-chiral fermion model defined
on the half line. The resonant point-contact tunneling interaction is transcribed into a non-local effective boundary interaction in the folded setup, where the boundary state formulation is applicable. 
We construct the boundary states for the models with two and three leads explicitly and evaluate the 
correlation functions of currents operators exactly. The electron transport between the leads is 
dominated by the resonant point-contact tunneling in the low frequency regime. 
We observe some $SU(2)$ and $SU(3)$ group theoretical
structures, which may be useful to analyze more complex models. 



\vskip 2cm

\section{INTRODUCTION}

In a recent paper we studied a model of resonant point-contact tunneling of a single lead \cite{Leechiral} by 
using the boundary state formulation. We have shown that the correlation functions of current operators are 
exactly calculable and the model exhibits the essential features of the models of this type 
\cite{Hewson,Saleur1998,Saleur2000,Affleck:1990by, Nayak} 
in the simplest form. In the present paper we shall discuss its extensions, which have multileads, by applying the same 
folding procedure and the boundary state formulation. Here we only discuss the Tomonaga-Luttinger (TL) liquids 
\cite{Tomonaga, Luttinger, Haldane} on the leads
at the critical point, so that the fermions on the leads are described by free fermion fields. In the folded setup the resonant point-contact tunneling between the leads is transcribed into a non-local boundary interaction. The main focus of the present work is the electron transport between the leads through the resonant point-contact tunneling. For this purpose we explicit evaluate the correlation functions of the 
current operators in the cases of models with two and three leads. The fermion field on each lead is labeled 
with the flavor indices. Thus, in the folded setup, we have free $SU(2)$ and $SU(3)$ Dirac fermion field actions as the bulk actions for the models with two and three leads respectively. The non-local boundary interactions break these global symmetries. 

The models we are about to explore in some detail may be considered as a variant of the multichannel Kondo model
\cite{Nozier80, Tsvelick85, Schlottmann, Affleck1993ld, Ludwig94, Kimcox}, yet in a simpler form. For more general discussions of its relations to the multichannel Kondo models and others, such as the quantum Brownian motion of a particle on lattices, the reader may refer to 
ref.\cite{Nayak}. The models are initially defined on an infinite line with a resonant contact 
interaction in the middle in terms of chiral fermion fields. If we divide the infinite line into two half lines,
and label the chiral fermion fields on each half line as the left and right movers, we can 
manufacture a two component Dirac fermion on a half line with a boundary interaction by folding. 
So the model perfectly fits into the boundary state formulation \cite{callan90, callan91, Callan1994, Lee:2005ge,
Hassel, Lee:06, Lee2008, Lee2009, Lee2009q, LeeU(1), LeeTL2015} of the closed string theory. Details of the 
folding procedure are given in the preceding work \cite{Leechiral}. 
In the present work for the models with multileads, we 
apply the folding procedure to each lead separately, labeling the Dirac fermion field on each lead with 
flavor indices. The boundary interactions, which are non-local and bilinears of fermion fields, are not 
diagonal in the flavor basis. We may diagonalize the boundary interactions by choosing a new basis, which 
is related to the flavor basis by a $SO(N)$ rotation transformation for the model with $N$ leads. 
Constructing the boundary state in the new basis leads us to the exact calculation of the 
correlation functions of the current operators. 

The rest of this paper is organized as follows: Section \ref{secn2} deals with the resonant point-contact 
tunneling model with two leads. Beginning with the model in the unfolded setup, we obtain the model in the 
folded setup. Diagonalizing the boundary interaction and the construction of the boundary state follow from the action of the model in the folded setup. Section \ref{secn3} is devoted to the resonant point-contact 
tunneling model with three leads. We repeat the same procedure, applied to the model with two leads, for the 
model with three leads. Some group theoretical structure is observed in the correlation functions of the 
current operators. Section \ref{secconclusions} concludes the paper with discussions on the exact results 
we obtain and an extension of the present work to the resonant point-contact tunneling model with 
four leads.

\section{Resonant Point-Contact Tunneling \\
Model with Two Leads} \label{secn2}

At a critical point the one-dimensional model for the point-contact tunneling with two
Tomonaga-Luttinger liquid leads may be described 
by the following action of two chiral fermion fields \cite{Nayak}
\beq
S 
= \frac{1}{2\pi}\int^\infty_{-\infty} d\s \int d\t \Bigl\{\sum_{a=1}^2 \eta^{a\dag} \left(\p_\t + i\p_\s \right) \eta^a
+ \delta(\s) d^{\dag} \p_\t d  +it \delta(\t) \sqrt{2}
\sum_{a=1}^2\left(\eta^{a\dag} d+ \eta^a d^\dag \right) \Bigr\}. \label{unfoldn2}
\eeq
This model may depict a resonant tunneling junction between two quantum wires. The chiral fermion fields 
$\eta^a$, $a=1, 2$ are labeled with flavor indices. The degrees of freedom of the impurity, located at the 
origin are denoted by the Fermi fields $d$ and $d^\dag$, which are the creation and annihilation operators of charge on the resonant state. We assume that the impurity does not have an internal structure for simplicity.
Since we are working on the theory at finite temperature, the theory is defined on a two dimensional Euclidean space and all Fermi fields are anti-periodic in $\t$. The physical parameters are scaled appropriately 
so that the range of $\t$ is $[0,2\pi]$. Since the TL liquids on the leads are at a critical point, 
their actions are just those of the free fermions and the coupling constant $t$ does not get renormalized. 
At a non-critical point $t$ may be replaced by a renormalized coupling constant. 

\subsection{The Model with Two Leads in the Unfolded Setup}

We shall begin with the model in the unfolded setup and transcribe it into the model in the folded setup 
\cite{fendley1995a, fendley1995, wong, Affleck1994} .
The equations of motion for the fermion fields $\eta^a$ in the unfolded setup follow from the action
Eq.(\ref{unfoldn2}) 
\begin{subequations}
\beq
(\p_\t + i\p_\s) \eta^a + it\sqrt{2}\delta(\s) d &=& 0, \label{eqn2a}\\
\p_\t d -it\sqrt{2} \sum_a \eta^a \bigl\vert_{\s =0} &=& 0, ~~~ a= 1, 2. \label{eqn2b}
\eeq
\end{subequations}
Since the fermi fields are anti-periodic in $\t$ we may write
\beq
\eta^a(\t,\s) = \sum_n \eta^a_n(\s) e^{in\t}, ~~ d(\t) = \sum_n d_n e^{in\t}, ~~ n \in {\bf Z} + 1/2 .
\eeq
If we rewrite the equations of motion Eqs.(\ref{eqn2a}, \ref{eqn2b}) in terms of normal modes, we have
\begin{subequations}
\beq
n \eta^a_n  &=& -  \p_\s \eta^a_n  -t \delta(\s) d_n , \label{first1}\\
n d_n  
&=& \frac{t}{\sqrt{2}} \sum_a
\left[\eta^a_n(0+)+ \eta^a_n(0-)\right] \label{second1}.
\eeq
\end{subequations}
From the first equation of motion, Eq.(\ref{first1}) we get the continuity condition as follows
\beq
\left[\eta^a_n(0+) - \eta^a_n(0-)\right] + t\sqrt{2}d_n &=& 0. \label{normal3}
\eeq
Then it follows from Eqs.(\ref{second1},\ref{normal3}) that
\beq
\left(I + \frac{t^2}{n} \bolM \right) \boleta_n(0+) = \left(I - \frac{t^2}{n} \bolM \right) \boleta_n(0-)\label{contn2}
\eeq 
where $I$ is an identity matrix and
\beq
\bolM = \begin{pmatrix} 1 & 1 \\ 1 & 1 \end{pmatrix}, ~~~ \boleta_n = \begin{pmatrix} \eta^1_n \\ \eta^2_n 
\end{pmatrix} .
\eeq
We may diagonalize the boundary conditions Eq.(\ref{contn2}), 
introducing a new basis for the Fermi fields, $\Psi^a_n$, $a= 1, 2$, which are defined as 
\beq
\Psi^1_n = \frac{1}{\sqrt{2}}\left(\eta^1_n + \eta^2_n \right), ~~
\Psi^2_n = \frac{1}{\sqrt{2}}\left(\eta^1_n - \eta^2_n \right) .
\eeq
The boundary conditions are now read as 
\beq
\Psi^1_n(0+) = \left(\frac{1- \frac{2t^2}{n}}{1+\frac{2t^2}{n}}\right) \Psi^1_n(0-), ~~
\Psi^2_n(0+) = \Psi^2_n(0-) . \label{boundunfold}
\eeq
As $t$ increases from zero, the boundary condition for $\Psi^1$ interpolates between the Neumann 
condition and the Dirichlet condition while the boundary condition for $\Psi^2$ remain same as the 
Neumann condition.

\subsection{The Model with Two Leads in the Folded Setup}

Utilizing the folding procedure discussed in the preceding study \cite{Leechiral} on the model 
with a single lead, 
we can easily construct the action for the model with two leads in the folded setup as
\beq
S_E &=& \frac{1}{2\pi}\int^\infty_{0} d\t \int d\s \left[ \barPsi^a \g \cdot \p \,\Psi^a \right] 
\nn\\
&&+ \frac{1}{2\pi} \int d\s \left\{ \barxi \g^1 \p_\s \xi + i\sqrt{2} t \left(
\barPsi^1 \g^1 \xi - \barxi \g^1 \Psi^1 \right) \right\}\Bigl\vert_{\t=0},
\eeq
where $\g^0 = \s_1$, $\g^1 = \s_2$, $\g^5 = -i \g^0 \g^1 = \s_3$. 
Note that we interchange the coordinates $\t$ and $\s$ in order to recast the model into the closed 
string picture, where the boundary state formulation is applicable. 
The chiral Fermi fields on each half line are combined together into two component Dirac spinors,
$\Psi^a$, $a=1, 2$, 
\begin{subequations}
\beq
\Psi^1 &=& \begin{pmatrix} \Psi^1_L  \\ \Psi^1_R \end{pmatrix}= \frac{1}{\sqrt{2}} 
\begin{pmatrix} \psi^1_L+\psi^2_L \\ \psi^1_R+\psi^2_R \end{pmatrix}, \\
\Psi^2 &=& \begin{pmatrix} \Psi^2_L  \\ \Psi^2_R \end{pmatrix}=\frac{1}{\sqrt{2}} 
\begin{pmatrix} \psi^1_L-\psi^2_L \\ \psi^1_R-\psi^2_R \end{pmatrix}.
\eeq
\end{subequations}
The Fermi field describing the degrees of freedom of the impurity is also extended to 
a two component Dirac spinor $\xi$.

\begin{figure}[htbp]
   \begin {center}
    \epsfxsize=0.6\hsize
%
	\epsfbox{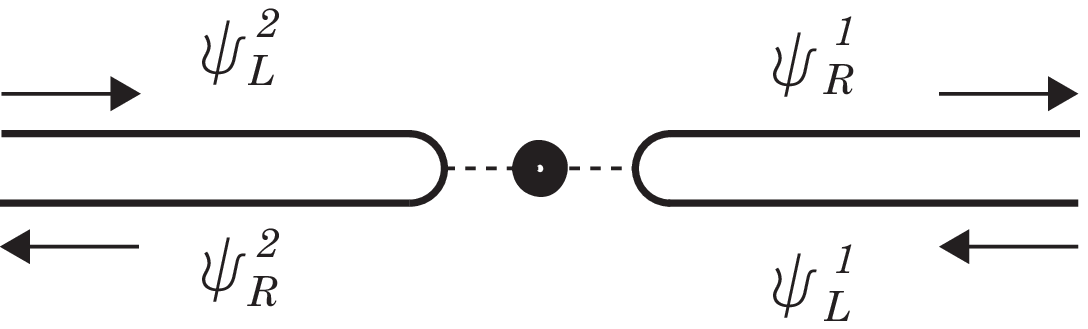}
   \end {center}
   \caption {\label{twoleads} Resonant Point-Contact Tunneling
with Two Leads in the Folded Setup}
\end{figure}

As we integrate out the fermi fields $\barxi $ and $\xi$, we obtain an effective non-local boundary action 
for $\Psi^1$
\beq
S_{\rm Boundary}\left[\barPsi^1, \Psi^1 \right] = 2t^2 \int \frac{d\s}{2\pi} \barPsi^1 \g^1 \left[\p_\s\right]^{-1}
\Psi^1 \Bigl\vert_{\t =0} .
\eeq
Accordingly the boundary state can be formally written as
\beq
\vert B \rangle = \exp \left\{2t^2 \int \frac{d\s}{2\pi} \barPsi^1 \g^1 \left[\p_\s\right]^{-1}
\Psi^1 \Bigl\vert_{\t =0} \right\}
\vert N, N \rangle. \label{formal}
\eeq
where the state $\vert N, N \rangle$ satisfies the Neumann boundary condition for both fermi fields,
$\Psi^1$ and $\Psi^2$.

\subsection{Boundary State for the Model with Two Leads}

It is straightforward to find an explicit expression of the boundary state $\vert B\rangle$
if we expand the two spinor components of the Fermi fields, $\Psi^a_L$ and $\Psi^a_R$ 
with the Fermion oscillators
\begin{subequations}
 \label{generallabel} \begin{eqnarray} \Psi^a_L(\tau+i\sigma)= \sum_n
 \Psi^a_n e^{-n(\tau+i\sigma)} ~~,~~
 \Psi^{a\dagger}_L(\tau+i\sigma)=\sum_n
 \Psi_n^{a\dagger}e^{-n(\tau+i\sigma)} \label{fermode:a}\\
 \Psi^a_R(\tau-i\sigma)=\sum_n \tilde \Psi^a_n e^{-n(\tau-i\sigma)} ~~,~~
 \Psi^{a\dagger}_R(\tau-i\sigma) = \sum_n \tilde \Psi_n^{a\dagger}
 e^{-n(\tau-i\sigma)}.\label{fermode:b} \end{eqnarray}
 \end{subequations} 
Since the Fermi fields $\Psi^a_{L/R}$ are anti-periodic,  
the oscillators $\Psi^a_n$ are labeled
by half-odd-integers, $n\in {\bf Z} +1/2$.
The non-vanishing anticommutation relations between the Fermion operators are 
 \begin{equation} \label{anticomns} \left\{\Psi^a_m, \Psi_n^{b\dagger}
 \right\} = \delta_{ab}\delta_{m+n} ~~~,~~~ \left\{ \tilde \Psi^a_m , \tilde
 \Psi_n^{b\dagger}\right\}=\delta_{ab}\delta_{m+n}. 
 \end{equation} 
The vacuum state $\vert 0 \rangle$ 
is annihilated by all positively moded oscillators 
\begin{equation}
\left.\begin{matrix} \Psi^a_n \left| 0\right\rangle=0,\,\, & \tilde
\Psi^a_n\left|0\right\rangle=0 \cr \Psi^{a\dagger}_n \left|
0\right\rangle=0,\,\, & \tilde \Psi^{a\dagger}_n\left|0\right\rangle=0
\cr \end{matrix} \right\}~~n>0 ,\end{equation}
and the Neumann state $\vert N, N \rangle$ satisfies 
\begin{subequations}
\beq
\Psi^a_n \vert N \rangle &=& i \tilde\Psi^{a\dag}_{-n} \vert N\rangle,~~
\Psi^{a\dag}_n \vert N \rangle = i \tilde \Psi^a_{-n} \vert N\rangle, \label{normalaa} \\
\tilde\Psi^a_n \vert N\rangle &=& -i \Psi^{a\dag}_{-n} \vert N \rangle,~~
\tilde\Psi^{a\dag}_n \vert N \rangle = -i \Psi^a_{-n} \vert N\rangle  \label{normalbb}
\eeq
\end{subequations}
where $n$ is a positive half-odd-integer; $n \in {\bf Z}+ 1/2$, $n >0$.
A solution of these equations is
\beq
 \left| N, N\right\rangle =  :\exp\left\{\sum_{a=1}^2 \sum_{n=1/2}^\infty i
 \left(\Psi_{-n}^{a\dagger}\tilde \Psi^{a\dagger}_{-n}+ \Psi^a_{-n}\tilde
 \Psi^a_{-n} \right)\right\}: \vert 0 \rangle.
\eeq

An explicit expression of the  boundary state $\vert B \rangle$ Eq.(\ref{formal}) in terms of 
the Fermion oscillators is found as
\beq \label{explicit}
\vert B\rangle 
&=& \prod_{n=1/2} :\exp \left\{-\frac{2t^2}{n} \tilde\Psi^{1\dag}_{-n} \tilde\Psi^1_n \right\}:
:\exp \left\{\frac{2t^2}{n} \tilde\Psi^{1\dag}_{n} \tilde\Psi^1_{-n} \right\}:\nn\\
&& 
:\exp\left\{-\frac{2t^2}{n}\Psi^{1\dag}_{-n}\Psi^1_n\right\} :
:\exp\left\{\frac{2t^2}{n}\Psi^{1\dag}_{n}\Psi^1_{-n}\right\} :
\vert N, N \rangle .
\eeq
It can be also written as 
\beq
\vert B \rangle &=& \prod_{n=1/2}^\infty:\exp\left\{-\frac{2t^2}{n} \Psi^{1\dag}_{-n} \Psi^1_n \right\} 
\exp \left\{-\frac{2t^2}{n} \tilde\Psi^1_{-n} \tilde\Psi_n^{1\dag} \right\}: \nn\\
&&
\exp\left\{i\left(1-\frac{2t^2}{n}\right) \Psi^{1\dag}_{-n} \tilde\Psi^{1\dag}_{-n}
\right\} \exp\left\{i\left(1-\frac{2t^2}{n}\right) \Psi^1_{-n} \tilde\Psi^1_{-n} \right\}
\nn \\
&& 
\exp\left\{i\Psi_{-n}^{2\dagger}\tilde \Psi^{2\dagger}_{-n}+ i\Psi^2_{-n}\tilde
\Psi^2_{-n} \right\} \vert 0 \rangle . \label{boundstaten2}
\eeq
By some algebra \cite{Leechiral}, we can show that the constructed boundary state satisfies the 
boundary conditions for $n =1/2, \dots $
\begin{subequations}
\beq
\Psi^1_n \vert B\rangle &=& i \left(\frac{1 - 2t^2/n}{1+ 2t^2/n}\right) \tilde\Psi^{1\dag}_{-n} \vert B \rangle, \label{boun2a} \\
\tilde\Psi^1_n \vert B\rangle &=& -i \left(\frac{1 - 2t^2/n}{1+ 2t^2/n}\right) \Psi^{1\dag}_{-n} \vert B \rangle,\label{boun2b} \\
\Psi^{1\dag}_n \vert B\rangle &=& i \left(\frac{1 - 2t^2/n}{1+ 2t^2/n}\right) \tilde\Psi^{1}_{-n} \vert B,
\label{boun2c}\\
\tilde\Psi^{1\dag}_n \vert B\rangle &=&-i \left(\frac{1 - 2t^2/n}{1+ 2t^2/n}\right) \Psi^{1}_{-n} 
\vert B\rangle . \label{boun2d}
\eeq
\end{subequations}
These conditions correspond to the boundary (continuity) conditions in the unfolded setup Eq.(\ref{boundunfold}). 
The right hand sides of Eqs.(\ref{boun2a}, \ref{boun2b}, \ref{boun2c}, \ref{boun2d}) differ from that of Eq.(\ref{boundunfold}) by a 
phase factor $\pm i$ due to the phase shift of the Fermi fields, which we adopt to transcribe the model in the 
unfolded setup into the model in the folded setup \cite{Leechiral}. 

The boundary conditions for $\Psi^2$ and $\tilde\Psi^2$ are just the Neumann condition, 
\begin{subequations}
\beq
\Psi^2_n \vert B\rangle &=& i  \tilde\Psi^{2\dag}_{-n} \vert B \rangle, \label{boun2e} \\
 \tilde\Psi^2_n\vert B\rangle &=& -i  \Psi^{2\dag}_{-n} \vert B \rangle,\label{boun2f} \\
\Psi^{2\dag}_n \vert B\rangle &=& i  \tilde\Psi^{2}_{-n} \vert B,\label{boun2g}\\
\tilde\Psi^{2\dag}_n \vert B\rangle &=&-i  \Psi^{2}_{-n} 
\vert B\rangle .\label{boun2h}
\eeq
\end{subequations}

\subsection{Correlation Functions of Current Operators of \\the Model with Two Leads}

The boundary state formulation is one of the most efficient methods to evaluate correlation functions of 
operators. The correlation functions of the operators $\calO_i$, $i = 1, \dots, n$, are calculated 
in the boundary state formulation as 
\beq
\langle T \calO_1 \dots \calO_n \rangle &=& \langle 0 \vert :  \calO_1 \dots \calO_n: \vert B \rangle/
\langle 0 \vert B \rangle . \label{correlation}
\eeq
Here we are interest in calculation of the correlation functions of the current operators, in particular,
\begin{subequations}
\beq
J^1_{L/R} &=& \psi^{1\dag}_{L/R} \psi^1_{L/R} = \frac{1}{2} \left(\Psi^{1\dag}_{L/R} + \Psi^{2\dag}_{L/R} \right) \left(\Psi^{1}_{L/R} + \Psi^{2}_{L/R} \right), \\
J^2_{L/R} &=& \psi^{2\dag}_{L/R} \psi^2_{L/R} =\frac{1}{2} \left(\Psi^{1\dag}_{L/R} - \Psi^{2\dag}_{L/R} \right) \left(\Psi^{1}_{L/R} - \Psi^{2}_{L/R} \right),
\eeq
\end{subequations}
from which the electron transport between the leads can be deduced. Since the boundary interaction 
is diagonal in the basis of ${\Psi^a, a =1, 2, 3}$, it may be convenient to write the current operators 
in this basis
 \begin{subequations}
\beq
J^1_L &=& \frac{1}{2} \bolPsi^\dag_L \left(I+\s^1 \right) \bolPsi_L, ~~
J^2_L = \frac{1}{2} \bolPsi^\dag_L \left(I-\s^1 \right) \bolPsi_L, \label{currentn2a} \\
J^1_R &=& \frac{1}{2} \bolPsi^\dag_R \left(I+\s^1 \right) \bolPsi_R, ~~
J^2_R = \frac{1}{2} \bolPsi^\dag_R \left(I-\s^1 \right) \bolPsi_R , \label{currentn2b}
\eeq
\end{subequations}
where
\beq
\bolPsi_L = \begin{pmatrix} \Psi^1_L \\ \Psi^2_L \end{pmatrix}, ~~~
\bolPsi_R = \begin{pmatrix} \Psi^1_R \\ \Psi^2_R \end{pmatrix}.
\eeq

Making use of the calculation of the correlation functions of the current operators for the 
case of a single lead \cite{Leechiral} and the representations of the current operators 
of the model with two leads Eqs.(\ref{currentn2a}, \ref{currentn2b}), we are able to calculate the 
correlation functions of the current operators. In what follows we only present the 
results of the calculations. The reader may refer to the preceding work \cite{Leechiral} for some 
details of techniques. We expect that the correlation functions of $\langle J^i_L (\t,\s) J^j_L(\tau^\prime,\sp)\rangle$ and $\langle J^i_R (\t,\s) J^j_R(\tau^\prime,\sp)\rangle$, $i, j = 1, 2$, are not 
affected by the boundary interaction. We may confirm it by an explicit calculation
\begin{subequations}
\beq
\langle 0 \vert J^1_L(\t,\s)  J^1_L(\tp, \s^\prime) \vert B\rangle
&=& \frac{1}{4} \left[2\frac{z z^\prime}{(z-z^\prime)^2}+2\frac{z z^\prime}{(z-z^\prime)^2}\right] \nn\\
&=& \frac{z z^\prime}{(z-z^\prime)^2}, \\
\langle 0 \vert J^1_L(\t,\s)  J^2_L(\tp, \s^\prime) \vert B\rangle
&=& \frac{1}{4} \left[2\frac{z z^\prime}{(z-z^\prime)^2}-2\frac{z z^\prime}{(z-z^\prime)^2}\right] \nn\\
&=& 0 , \\
\langle 0 \vert J^2_L(\t, \s)  J^1_L(\tp, \s^\prime) \vert B\rangle 
&=& 0,\\
\langle 0 \vert J^2_L(\t, \s)  J^2_L(\tp, \s^\prime) \vert B\rangle 
&=&\frac{z z^\prime}{(z-z^\prime)^2},
\eeq 
\end{subequations}
where $z = e^{\t+i\s}$, $\zp =e^{\tp+i\sp}$. 
The correlation functions in the right moving sector are evaluated similarly
\begin{subequations}
\beq
\langle 0 \vert J^1_R(\t, \s)  J^1_R(\tp, \s^\prime) \vert B\rangle &=& \frac{\barz \barz^\prime}{(\barz-\barz^\prime)^2}, \\
\langle 0 \vert J^1_R(\t, \s)  J^2_R(\tp, \s^\prime) \vert B\rangle &=& 0, \\
\langle 0 \vert J^2_R(\t, \s)  J^1_R(\tp, \s^\prime) \vert B\rangle &=& 0, \\
\langle 0 \vert J^2_R(\t, \s)  J^2_R(\tp, \s^\prime) \vert B\rangle &=& \frac{\barz \barz^\prime}{(\barz-\barz^\prime)^2}, 
\eeq
\end{subequations}
where $\barz = e^{\t-i\s}$, $\barz^\prime =e^{\tp-i\sp}$.

The nontrivial correlation functions are those of $\langle J^i_L (\t,\s) J^j_R(\tau^\prime,\sp)\rangle$ and $\langle J^i_R (\t,\s) J^j_L(\tau^\prime,\sp)\rangle$, $i, j = 1, 2$, which depict the electron transport between the leads 
by the resonant point-contact tunneling,
\begin{subequations}
\beq
\langle 0 \vert J^1_L(\t, \s)  J^1_R(\tp, \s^\prime) \vert B\rangle 
&=& \frac{z\barz^\prime}{\left(z\barz^\prime -1\right)^2} -3t^2 \frac{\sqrt{z\barz^\prime}}{z\barz^\prime -1}
\ln \frac{\sqrt{z\barz^\prime} +1}{\sqrt{z\barz^\prime} -1} \nn\\
&&~~~~~~~~~~~~~~+ 4t^4 \left(\ln \frac{\sqrt{z\barz^\prime} +1}{\sqrt{z\barz^\prime} -1}\right)^2, \\
\langle 0 \vert J^1_L(\t, \s)  J^2_R(\tp, \s^\prime) \vert B\rangle 
&=& -t^2 \frac{\sqrt{z\barz^\prime}}{z\barz^\prime -1}
\ln \frac{\sqrt{z\barz^\prime} +1}{\sqrt{z\barz^\prime} -1} + 4t^4 \left(\ln \frac{\sqrt{z\barz^\prime} +1}{\sqrt{z\barz^\prime} -1}\right)^2, \\
\langle 0 \vert J^2_L(\t, \s)  J^2_R(\tp, \s^\prime) \vert B\rangle &=& \langle 0 \vert J^1_L(\s)  J^1_R(\s^\prime) \vert B\rangle, \\
\langle 0 \vert J^2_L(\t, \s)  J^1_R(\tp, \s^\prime) \vert B\rangle &=& -t^2 \frac{\sqrt{\barz z^\prime}}{\barz z^\prime -1}
\ln \frac{\sqrt{\barz z^\prime} +1}{\sqrt{\barz z^\prime} -1} + 4t^4 \left(\ln \frac{\sqrt{\barz z^\prime} +1}{\sqrt{\barz z^\prime} -1}\right)^2.
\eeq 
\end{subequations}
Note that the electron transport between different leads is solely due to the tunneling interaction, which is enhanced in the low frequency regime. As we can see 
that the contributions of the tunneling interaction is long range and insensitive to the high frequency modes due to the non-local nature of the contact tunneling interaction. We may deduce the frequency dependent conductance from the correlation functions of the current operators.

\section{Resonant Point-Contact Tunneling \\
Model with Three Leads} \label{secn3}

Now we are are ready to discuss the resonant point-contact tunneling model with three leads, which is more 
complex, yet more interesting. The model may be also relevant to the study of the junction of the three quantum wires or the $Y$-shape junction \cite{chamon, oshikawa2006, Giuliano08, aristov2011, aristov2013, Shi2016}. 

\subsection{The Model with Three Leads in the Unfolded Setup}

The resonant point-contact tunneling model with three leads is described in the unfolded setup by the same action Eq.(\ref{unfoldn2}) of the model
with two leads, except that three chiral fermion fields $\eta^a$, $a=1, 2, 3$, are introduced
\beq
S 
&=& \frac{1}{2\pi}\int^\infty_{-\infty} d\s \int d\t \Bigl\{\sum_{a=1}^3 \eta^{a\dag} \left(\p_\t + i\p_\s \right) \eta^a \nn\\
&& ~~~~~~
+ \delta(\s) d^{\dag} \p_\t d  +it \delta(\t) \sqrt{2}
\sum_{a=1}^3\left(\eta^{a\dag} d+ \eta^a d^\dag \right) \Bigr\}. \label{unfoldn3}
\eeq
The continuity condition (boundary condition) for the model with three leads is obtained as 
\begin{subequations}
\beq
\left(I + \frac{t^2}{n} \bolM \right) \boleta_n(0+) &=& \left(I - \frac{t^2}{n} \bolM \right) \boleta_n(0-),
\label{contin3} \\
\bolM &=& \begin{pmatrix} 1 & 1 & 1\\ 1 & 1 & 1\\1 & 1 & 1 \end{pmatrix}, ~~~
\boleta_n = \begin{pmatrix} \eta^1_n \\ \eta^2_n \\ \eta^3_n \end{pmatrix}. \label{m3}
\eeq
\end{subequations}
The matrix $\bolM$, Eq.(\ref{m3}) has a set of orthonormal eigenvectors,
\beq
\frac{1}{\sqrt{3}} \begin{pmatrix} 1 \\  1 \\ 1 \end{pmatrix}, ~~ 
\frac{1}{\sqrt{2}} \begin{pmatrix} ~~1 \\  -1 \\ ~~0 \end{pmatrix}, ~~ 
\frac{1}{\sqrt{6}} \begin{pmatrix} ~~1 \\  ~~1 \\ -2 \end{pmatrix},
\eeq
of which eigenvalues are $\{3, 0, 0 \} $ respectively.

We may transform the matrix $\bolM$ to a diagonal one by an $SO(3)$ rotation 
\beq
\bolR^t \bolM \bolR = \begin{pmatrix} 3 & 0 & 0 \\ 0 & 0 & 0 \\ 0 & 0 & 0 \end{pmatrix} ,
\eeq
where 
\beq
{\boldsymbol R} =\begin{pmatrix} \frac{1}{\sqrt{3}} & \frac{1}{\sqrt{2}} & \frac{1}{\sqrt{6}} \\
\frac{1}{\sqrt{3}} & -\frac{1}{\sqrt{2}} & \frac{1}{\sqrt{6}} \\
\frac{1}{\sqrt{3}} & 0 & -\frac{\sqrt{2}}{\sqrt{3}} \end{pmatrix}, ~~~
{\boldsymbol R}^t {\boldsymbol R} = 
{\boldsymbol R}{\boldsymbol R}^t= I . 
\eeq

The continuity condition Eqs.(\ref{contin3}, \ref{m3}) is diagonalized as we introduce 
a new basis for the Fermi fields, $\Psi^a$ which are
related to the chiral Fermi fields $\eta^a$ as follows 
\beq
\Psi^a = (\bolR^t)^a{}_b \, \eta^b, ~~~ a, b = 1, 2, 3. 
\eeq
To be explicit,
\begin{subequations}
\beq
\Psi^1 &=& \frac{1}{\sqrt{3}}\eta^1 + \frac{1}{\sqrt{3}}\eta^2 + \frac{1}{\sqrt{3}} \eta^3, \\
\Psi^2 &=& \frac{1}{\sqrt{2}}\eta^1  -\frac{1}{\sqrt{2}}\eta^2, \\
\Psi^3 &=& \frac{1}{\sqrt{6}}\eta^1 + \frac{1}{\sqrt{6}}\eta^2  -\frac{\sqrt{2}}{\sqrt{3}}\eta^3 .
\eeq
\end{subequations}
The continuity conditions, which may be also understood as boundary conditions are read as follows
\beq
\Psi^1_n(0+) = \left(\frac{1- \frac{3t^2}{n}}{1+\frac{3t^2}{n}}\right) \Psi^1_n(0-), ~~
\Psi^2_n(0+) = \Psi^2_n(0-), ~~ \Psi^3_n(0+) = \Psi^3_n(0-).
\eeq
As the coupling constant $t$ varies, the boundary condition for $\Psi^1$ interpolates from the Neumann
condition to the Dirichlet condition. 

\subsection{The Model with Three Leads in the Folded Setup} 

From our study on the model with a single and two leads, we can easily construct the action for the model
with three leads in the folded setup as
\beq
S_E &=& \frac{1}{2\pi}\int^\infty_{0} d\t \int d\s \left[ \sum_{a=1}^3\barPsi^a \g \cdot \p \,\Psi^a \right] 
\nn\\
&&+ \frac{1}{2\pi} \int_{\t=0} d\s \left\{ \barxi \g^1 \p_\s \xi + i\sqrt{3} t \left(
\barPsi^1 \g^1 \xi - \barxi \g^1 \Psi^1 \right) \right\},
\eeq
where
\begin{subequations}
\beq
\Psi^1 &=& \begin{pmatrix} \Psi^1_L  \\ \Psi^1_R \end{pmatrix}
= \frac{1}{\sqrt{3}}\begin{pmatrix} \psi^1_L+\psi^2_L+ \psi^3_L \\\psi^1_R+\psi^2_R+ \psi^3_R
\end{pmatrix}, \\
\Psi^2 &=& \begin{pmatrix} \Psi^2_L  \\ \Psi^2_R \end{pmatrix}
=\frac{1}{\sqrt{2}} \begin{pmatrix} \psi^1_L-\psi^2_L \\  \psi^1_R-\psi^2_R \end{pmatrix},  \\
\Psi^3 &=& \begin{pmatrix} \Psi^1_L  \\ \Psi^1_R \end{pmatrix}
= \frac{1}{\sqrt{6}} \begin{pmatrix} \psi^1_L+\psi^2_L-2 \psi^3_L \\
\psi^1_R+\psi^2_R-2 \psi^3_R \end{pmatrix}.
\eeq
\end{subequations}
Integrating out the fermi fields $\barxi $ and $\xi$, we obtain the effective boundary action 
for $\barPsi^1$ and $\Psi^1$
\beq
S_{\rm eff}\left[\barPsi^1, \Psi^1 \right] = 3t^2 \int \frac{d\s}{2\pi} \barPsi^1 \g^1 \left[\p_\s\right]^{-1}
\Psi^1 \Bigl\vert_{\t =0} .
\eeq
Accordingly the boundary state can be written as
\beq
\vert B \rangle = \exp \left\{3t^2 \int \frac{d\s}{2\pi} \barPsi^1 \g^1 \left[\p_\s\right]^{-1}
\Psi^1 \Bigl\vert_{\t =0} \right\}
\vert N, N, N \rangle \label{bounstaten3}
\eeq
where the state $\vert N, N, N \rangle$ satisfies the Neumann boundary conditions for fermi fields,
$\Psi^1$, $\Psi^2$ and $\Psi^3$ . 

\begin{figure}[htbp]
   \begin {center}
    \epsfxsize=0.6\hsize
%
	\epsfbox{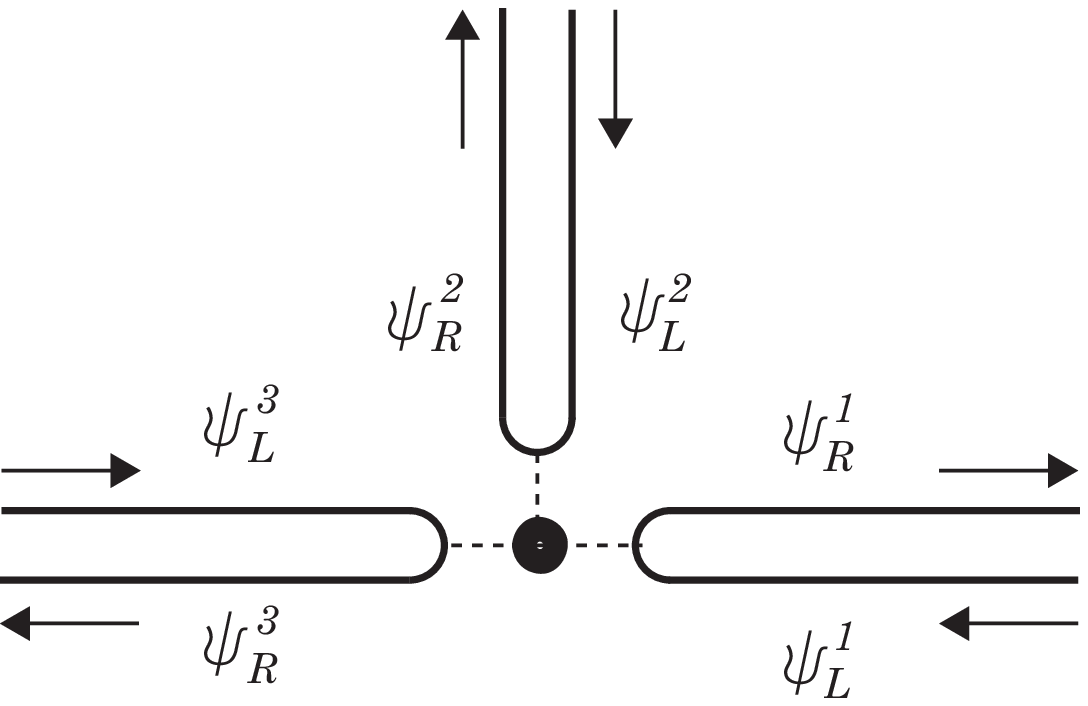}
   \end {center}
   \caption {\label{threeleads} Resonant Point-Contact Tunneling
with Three Leads in the Folded Setup}
\end{figure}

\subsection{Boundary State for the Model with Three Leads} 

The boundary conditions for $\Psi^1$ and $\tilde \Psi^1$ are 
\begin{subequations}
\beq
\Psi^1_n \vert B\rangle &=& i \left(\frac{1 - 3t^2/n}{1+ 3t^2/n}\right) \tilde\Psi^{1\dag}_{-n} \vert B \rangle, ~~ \tilde\Psi^1_n = -i \left(\frac{1 - 3t^2/n}{1+ 3t^2/n}\right) \Psi^{1\dag}_{-n} \vert B \rangle ,
\label{bounn3a}\\
\Psi^{1\dag}_n \vert B\rangle &=& i \left(\frac{1 - 3t^2/n}{1+ 3t^2/n}\right) \tilde\Psi^{1}_{-n} \vert B,~~
\tilde\Psi^{1\dag}_n \vert B\rangle =-i \left(\frac{1 - 3t^2/n}{1+ 3t^2/n}\right) \Psi^{1}_{-n} 
\vert B\rangle, \label{bounn3b}
\eeq
\end{subequations}
and the Fermi field operators $\Psi^a$ and $\tilde\Psi^a$,  $a=2, 3$ satisfy the Neumann condition, 
\begin{subequations}
\beq
\Psi^a_n \vert B\rangle &=& i  \tilde\Psi^{a\dag}_{-n} \vert B \rangle, ~~
 \tilde\Psi^a_n = -i  \Psi^{a\dag}_{-n} \vert B \rangle \label{bounn3c}\\
\Psi^{a\dag}_n \vert B\rangle &=& i  \tilde\Psi^{a}_{-n} \vert B,~~
\tilde\Psi^{a\dag}_n \vert B\rangle =-i  \Psi^{a}_{-n}\vert B\rangle,  ~~~a = 2, 3 . \label{bounn3d}
\eeq
\end{subequations}
These boundary conditions Eqs.(\ref{bounn3a},\ref{bounn3b},\ref{bounn3c},\ref{bounn3d}) are satisfied by the 
constructed boundary state Eq.(\ref{bounstaten3}), of which expression in terms of the normal modes is given as 
\beq
\vert B \rangle &=& \prod_{n=1/2}^\infty:\exp\left\{-\frac{3t^2}{n} \Psi^{1\dag}_{-n} \Psi^1_n \right\} 
\exp \left\{-\frac{3t^2}{n} \tilde\Psi^1_{-n} \tilde\Psi_n^{1\dag} \right\}: \nn\\
&&  
\exp\left\{i\left(1-\frac{3t^2}{n}\right) \Psi^{1\dag}_{-n} \tilde\Psi^{1\dag}_{-n}
\right\} \exp\left\{i\left(1-\frac{3t^2}{n}\right) \Psi^1_{-n} \tilde\Psi^1_{-n} \right\}
\nn \\
&& 
\prod_{a=2}^3:\exp\left\{i\Psi_{-n}^{a\dagger}\tilde \Psi^{a\dagger}_{-n}+ i\Psi^a_{-n}\tilde
 \Psi^a_{-n} \right\}: \vert 0 \rangle . \label{boundaryn3}
\eeq

\subsection{Correlation Functions of Currents Operators of \\
the Model with Three Leads}

The current operators on each lead may be written in terms of the Fermi fields of the basis
$\{\Psi^a, a=1, 2, 3 \}$ as 
\begin{subequations}
\beq
J^a_{L} &=& \psi^{a\dag}_{L} \psi^a_{L} = \sum_{b,c} \Psi^{b\dag}_L (\bolR^t)_{ba} (\bolR)_{ac}\Psi^c_L, \\
J^a_{R} &=& \psi^{a\dag}_{R} \psi^a_{R} = \sum_{b,c} \Psi^{b\dag}_R (\bolR^t)_{ba} (\bolR)_{ac}\Psi^c_R.
\eeq
\end{subequations}
It may be useful to spell out the explicit expression of the current operators for the purpose of calculating the correlation functions of the current operators
\begin{subequations}
\beq
J^1_{L/R} &=& \Psi^\dag_{L/R} \bolD_1 \Psi_{L/R},~~ (\bolD_1)_{ab} = (\bolR)_{1a} (\bolR)_{1b}, \\
J^2_{L/R} &=& \Psi^\dag_{L/R} \bolD_2 \Psi_{L/R},~~ (\bolD_2)_{ab} = (\bolR)_{2a} (\bolR)_{2b}, \\
J^3_{L/R} &=& \Psi^\dag_{L/R} \bolD_3 \Psi_{L/R},~~ (\bolD_3)_{ab} = (\bolR)_{3a} (\bolR)_{3b},
\eeq
\end{subequations}
where
\begin{subequations}
\beq
\bolD_1 &=& 
\begin{pmatrix} \frac{1}{3} & \frac{1}{\sqrt{6}} & \frac{1}{3\sqrt{2}} \\
\frac{1}{\sqrt{6}} & \frac{1}{2} &\frac{1}{2\sqrt{3}} \\
\frac{1}{3\sqrt{2}} & \frac{1}{2\sqrt{3}} & \frac{1}{6}
\end{pmatrix}, \\
\bolD_2 &=& 
\begin{pmatrix} \frac{1}{3} & -\frac{1}{\sqrt{6}} & \frac{1}{3\sqrt{2}} \\
-\frac{1}{\sqrt{6}} & \frac{1}{2} & -\frac{1}{2\sqrt{3}} \\
\frac{1}{3\sqrt{2}} & -\frac{1}{2\sqrt{3}} & \frac{1}{6}
\end{pmatrix}, \\ 
\bolD_3 &=& 
\begin{pmatrix} \frac{1}{3} & 0 & -\frac{\sqrt{2}}{3} \\
0  & 0 & 0 \\
-\frac{\sqrt{2}}{3} & 0  & \frac{2}{3}
\end{pmatrix}.
\eeq
\end{subequations}
These are symmetric real matrices, so that they can be represented in terms of the $SU(3)$ Gellmann matrices,
$\l^a$, $a =1, \dots, 8$, 
\begin{subequations}
\beq
\bolD_1 &=& \frac{1}{3} I + \frac{1}{\sqrt{6}} \l_1 - \frac{1}{12} \l_3 + \frac{1}{3\sqrt{2}} \l_4
+ \frac{1}{2\sqrt{3}} \l_6 + \frac{1}{4\sqrt{3}} \l_8 ,\\
\bolD_2 &=& \frac{1}{3} I - \frac{1}{\sqrt{6}} \l_1 - \frac{1}{12} \l_3 + \frac{1}{3\sqrt{2}} \l_4
- \frac{1}{2\sqrt{3}} \l_6 + \frac{1}{4\sqrt{3}} \l_8 , \\
\bolD_3 &=& \frac{1}{3} I + \frac{1}{6} \l_3- \frac{\sqrt{2}}{3} \l_4  - \frac{1}{2\sqrt{3}} \l_8 .
\eeq
\end{subequations}

We already know that the correlation functions of $\langle J^i_L(\t,\s) J^j_L(\tp, \sp)\rangle$ and 
$\langle J^i_R(\t,\s) J^j_R(\tp, \sp)\rangle$, $i, j = 1, 2, 3$, are not affected by the 
boundary interaction. But it may be worthwhile to confirm it by an explicit calculation:
\begin{subequations}
\beq
\langle 0 \vert J^1_L(\s) J^1_L(\sp) \vert B \rangle 
&=& \left[\frac{1}{9}+ \frac{1}{4}+ \frac{1}{36} + 2 \left(\frac{1}{6}+ \frac{1}{18} + \frac{1}{12} \right) \right] 
\frac{z z^\prime}{(z-z^\prime)^2} \nn\\
&=& \frac{z z^\prime}{(z-z^\prime)^2}, \\
\langle 0 \vert J^1_L(\s) J^2_L(\sp) \vert B \rangle
&=& \left[\frac{1}{9}+ \frac{1}{4}+ \frac{1}{36} -2 \left(\frac{1}{6}+  \frac{1}{12} \right) + \frac{2}{9\cdot 2} \right] \frac{z z^\prime}{(z-z^\prime)^2} \nn\\
&=& 0 . \\
\langle 0 \vert J^2_L(\s) J^2_L(\sp) \vert B \rangle
&=& \left[\frac{1}{9}+ \frac{1}{4}+ \frac{1}{36} + 2 \left(\frac{1}{6}+ \frac{1}{18} + \frac{1}{12} \right) \right] \nn\\ 
&=& \frac{z z^\prime}{(z-z^\prime)^2}, \\
\langle 0 \vert J^1_L(\s) J^3_L(\sp) \vert B \rangle
&=& \left[\frac{1}{9} + \frac{1}{6} \frac{2}{3} - 2 \frac{\sqrt{2}}{3} \frac{1}{3\sqrt{2}} \right]
\frac{z z^\prime}{(z-z^\prime)^2} \nn\\
&=& 0, \\
\langle 0 \vert J^2_L(\s) J^3_L(\sp) \vert B \rangle
&=& \left[\frac{1}{9} + \frac{1}{6} \frac{2}{3} - 2 \frac{\sqrt{2}}{3} \frac{1}{3\sqrt{2}} \right]
\frac{z z^\prime}{(z-z^\prime)^2} \nn\\
&=& 0 , \\
\langle 0 \vert J^3_L(\s) J^3_L(\sp) \vert B \rangle 
&=& \left[\frac{1}{9} + \frac{4}{9} + 2 \left(-\frac{\sqrt{2}}{3}\right)^2 \right] \frac{z z^\prime}{(z-z^\prime)^2} 
\nn\\
&=& \frac{z z^\prime}{(z-z^\prime)^2}. 
\eeq
\end{subequations}
These correlation functions are summarized as 
\beq
\langle 0 \vert J^i_L(\t,\s) J^j_L(\tp, \sp) \vert B \rangle &=& \delta^{ij} \frac{z \zp}{(z-\zp)^2} ,\\
i, j &=& 1, 2, 3. \nn 
\eeq
The correlation functions of the right moving sector can be also calculated similarly and 
are succinctly summarized as 
\beq
\langle 0 \vert J^i_R(\t,\s) J^j_R(\tp, \sp) \vert B \rangle &=& \delta^{ij} \frac{\barz \barzp}{(\barz-\barzp)^2} ,\\
i, j &=& 1, 2, 3. \nn 
\eeq

The non-trivial correlation functions of the current operators, affected by the tunneling interaction with
the impurity are $\langle J^i_L(\t,\s) J^j_R(\tp, \sp) \rangle$, $i, j = 1, 2, 3$: The results of 
the explicit calculation may be summarized as
\begin{subequations}
\beq
\langle 0 \vert J^i_L(\t, \s) J^j_R(\tp, \sp) \vert B \rangle 
&=& \delta_{ij} F(z, \barzp)+ \left(\bolM- I \right)_{ij} G(z, \barzp), \label{currentn3ij}\\
F(z, \barzp)&=& \frac{z\barz^\prime}{\left(z\barz^\prime -1\right)^2} - \frac{8}{3} t^2\frac{\sqrt{z\barz^\prime}}{z\barz^\prime -1}\ln \frac{\sqrt{z\barz^\prime} +1}{\sqrt{z\barz^\prime} -1} \nn\\
&& + 4t^4 \left(\ln \frac{\sqrt{z\barz^\prime} +1}{\sqrt{z\barz^\prime} -1}\right)^2 ,\\
G(z, \barzp) &=&  -\frac{2}{3} t^2\frac{\sqrt{z\barz^\prime}}{z\barz^\prime -1}\ln \frac{\sqrt{z\barz^\prime} +1}{\sqrt{z\barz^\prime} -1} + 4t^4 \left(\ln \frac{\sqrt{z\barz^\prime} +1}{\sqrt{z\barz^\prime} -1}\right)^2.
\eeq
\end{subequations}
The second term in Eq.(\ref{currentn3ij}) describes the electron transport between the leads due to the resonant 
tunneling, which is dominant in the low frequency regime. The resonant tunneling interaction with the impurity 
suppresses the reflection currents on the leads but enhances the electron transport between different leads
as the frequency is lowered.  

\section{Discussions and Conclusions} \label{secconclusions}

We conclude the paper with a brief discussion on the extension to the model with four leads and a remark on the 
boundary state. In this paper, 
we studied the multilead point-contact tunneling, which may be relevant to the multi-channel Kondo problem and
the junctions of quantum wires, using the boundary state formulation. Although we only dealt with the 
models with two and three leads, our discussion can be certainly extended to the model with more than three 
leads. We may discuss the application of the boundary state formulation to the model with four leads briefly.
The matrix $\bolM$, which enters the boundary conditions Eq.(\ref{m3}), is replaced by a $4 \times 4$ matrix 
\beq
\bolM = \begin{pmatrix} 1 & 1 & 1 & 1\\ 1 & 1 & 1 & 1\\1 & 1 & 1 & 1 \\1 & 1 & 1 & 1 \end{pmatrix}. 
\eeq
Its eigenvalues are $\{4, 0, 0, 0\}$ and the corresponding orthonormal eigenvectors are respectively 
\beq
\frac{1}{2}\begin{pmatrix} 1 \\ 1 \\ 1 \\ 1 \end{pmatrix}, ~~ 
\frac{1}{\sqrt{2}}\begin{pmatrix} ~~1 \\ -1 \\ ~~0 \\~~ 0 \end{pmatrix},  ~~ 
\frac{1}{\sqrt{6}}\begin{pmatrix} ~~1 \\  ~~1 \\-2 \\ ~~0 \end{pmatrix},~~
\frac{1}{2\sqrt{3}}\begin{pmatrix} ~~1 \\ ~~1 \\ ~~1 \\ -3 \end{pmatrix}. 
\eeq
An $SO(4)$ rotation transforms the matrix $\bolM$ into a diagonal matrix 
\beq
\bolR^t \bolM \bolR &=& \begin{pmatrix} 4 & 0 & 0 & 0 \\ 0 & 0 & 0 & 0 \\
0 & 0 & 0 & 0 \\ 0 & 0 & 0 & 0 \end{pmatrix}, 
\eeq
where
\beq
\bolR &=& \begin{pmatrix} \half & \frac{1}{\sqrt{2}} & \frac{1}{\sqrt{6}} & \frac{1}{2\sqrt{3}} \\
\half & -\frac{1}{\sqrt{2}} & \frac{1}{\sqrt{6}} & \frac{1}{2\sqrt{3}} \\
\half & 0 & - \sqrt{\frac{2}{3}} & \frac{1}{2\sqrt{3}} \\
\half & 0 & 0 & -\frac{\sqrt{3}}{2} \end{pmatrix}, ~~~\bolR^t \bolR = \bolR \bolR^t = I .
\eeq
The Fermi fields of the new basis $\Psi^a$, $a=1, 2, 3, 4$, by which the boundary interaction is 
diagonal are also related to the 
Fermi fields $\psi^a$, $a = 1, 2, 3, 4$, on each lead (lead basis) by the $SO(4)$ rotation
\beq
\Psi^a_{L/R} =\sum_{b=1}^4 (\bolR^t)_{ab} \psi^b_{L/R}. \label{so4rot}
\eeq
To be explicit, we have
\begin{subequations}
\beq
\Psi^1_{L/R} &=&  \half\psi^1_{L/R} + \half \psi^2_{L/R} + \half\psi^3_{L/R} + \half\psi^4_{L/R} , \\
\Psi^2_{L/R} &=&  \frac{1}{\sqrt{2}}\psi^1_{L/R} -\frac{1}{\sqrt{2}}\psi^2_{L/R}, \\
\Psi^3_{L/R} &=&  \frac{1}{\sqrt{6}}\psi^1_{L/R} + \frac{1}{\sqrt{6}}\psi^2_{L/R}  -\frac{\sqrt{2}}{\sqrt{3}}\psi^3_{L/R}, \\
\Psi^4_{L/R} &=&  \frac{1}{2\sqrt{3}}\psi^1_{L/R} + \frac{1}{2\sqrt{3}}\psi^2_{L/R} + \frac{1}{2\sqrt{3}}\psi^3_{L/R}  - \frac{\sqrt{3}}{2}\psi^4_{L/R} .
\eeq
\end{subequations}
The correlation functions of the current operators may follow from Eqs(\ref{so4rot},\ref{correlation}) and the boundary state for the model with four leads, which is given as 
\beq
\vert B \rangle &=& \prod_{n=1/2}^\infty:\exp\left\{-\frac{4t^2}{n} \Psi^{1\dag}_{-n} \Psi^1_n \right\} 
\exp \left\{-\frac{4t^2}{n} \tilde\Psi^1_{-n} \tilde\Psi_n^{1\dag} \right\}: \nn\\
&&  
\exp\left\{i\left(1-\frac{4t^2}{n}\right) \Psi^{1\dag}_{-n} \tilde\Psi^{1\dag}_{-n}
\right\} \exp\left\{i\left(1-\frac{4t^2}{n}\right) \Psi^1_{-n} \tilde\Psi^1_{-n} \right\}
\nn \\
&& 
\prod_{a=2}^4\exp\left\{i\Psi_{-n}^{a\dagger}\tilde \Psi^{a\dagger}_{-n}+ i\Psi^a_{-n}\tilde
 \Psi^a_{-n} \right\} \vert 0 \rangle. \label{bounn4}
\eeq
Since the evaluation of the correlation functions of the current operators is straightforward, we may leave it 
as an exercise for the reader, who might be interested in. 

Finally a remark on the boundary state concludes the paper.  
Following refs.\cite{callan90, Callan1994}, we defined the boundary state as 
\beq
\vert B \rangle = \exp\left(- S_{boundary} \right) \vert N \rangle . \label{def}
\eeq
It should be noted that the boundary interaction is treated as a perturbation. Hence the 
calculations based on the boundary states Eqs.(\ref{boundstaten2}, \ref{boundaryn3}, \ref{bounn4}) 
may be trusted only for small $t$. For large $t$,
we may employ alternative forms of the boundary states.  
For an example, we may replace the boundary state
Eq.(\ref{boundstaten2}) for the model with two leads by the following one
\beq
\vert B \rangle &=& \prod_{n=1/2}^\infty\exp\left\{i \left(\frac{1-2t^2/n}{1+2t^2/n}\right)
\left(\Psi_{-n}^{1\dagger}\tilde \Psi^{1\dagger}_{-n}+ \Psi^1_{-n}\tilde
\Psi^1_{-n}\right) \right\} \nn\\
&&~~~~~~
\exp\left\{i\left(\Psi_{-n}^{2\dagger}\tilde \Psi^{2\dagger}_{-n}+ \Psi^2_{-n}\tilde
\Psi^2_{-n}\right) \right\} \vert 0 \rangle . \label{altern2}
\eeq
This boundary state satisfies the same boundary conditions 
Eqs.(\ref{boun2a},\ref{boun2b},\ref{boun2c},\ref{boun2d}, \ref{boun2e}, \ref{boun2f}, \ref{boun2f}, \ref{boun2h}), although it cannot be put in the form of 
Eq.(\ref{def}). As $t$ increases from zero to infinity, the boundary state Eq.(\ref{altern2}) interpolates
between the Neumann state and the Dirichlet state. For small $t$, at the leading order both expressions of the boundary states Eq.(\ref{boundstaten2}) and Eq.(\ref{altern2}) agree. Further studies may be needed before 
employing the boundary state Eq.(\ref{altern2}) to study the resonant point-contact tunneling at intermediate and large 
values of the coupling constant.

\vskip 1cm

\noindent{\bf Acknowledgments}\\
This work was supported by Kangwon National University Grant 2013.


\begin{thebibliography}{0}

\bibitem{Leechiral}
T. Lee,
{\it Chiral Fermion and Boundary State Formulation: Resonant Point-Contact Tunneling},
to be published in IJMPB,
[arXiv:1509.03960] (2015).

\bibitem{Hewson} A. C. Hewson, 
{\it The Kondo Problem to Heavy Fermions}, 
Cambridge University Press (1997).

\bibitem{Saleur1998} H. Saleur, 
{\it Lectures on Non Perturbative Field Theory and Quantum Impurity Problems},
[arXiv:cond-mat/9812110v1] (1998).

\bibitem{Saleur2000} H. Saleur, 
{\it Lectures on Non Perturbative Field Theory and Quantum Impurity Problems: Part II},
[arXiv:cond-mat/0007309] (2000).

\bibitem{Affleck:1990by}
I.~Affleck and A.~W.~W.~Ludwig,
Nucl.\ Phys.\ B {\bf 352}, 849 (1991).

\bibitem{Nayak}  C. Nayak, M. P. A. Fisher, A. W. W. Ludwig and H. H. Lin, 
Phys. Rev. B {\bf 59}, 15 694 (1999).

\bibitem{Tomonaga} S. Tomogana, 
Prog. Theor. Phys. {\bf 5}, 544 (1950). 

\bibitem{Luttinger}
J. M. Luttinger, 
J. Math. Phys. {\bf 4}, 1154 (1963). 

\bibitem{Haldane} F. D. M. Haldane,
J. Phys. C. {\bf 14}, 2585 (1981).

\bibitem{Nozier80}
P. Nozieres and A. Blandin, 
J. Phys. (Paris) {\bf 41}, 193 (1980).

\bibitem{Tsvelick85}
A. M. Tsvelick and P. B. Wiegmann,
J. of Stat. Phys. {\bf 38}, 125 (1985).

\bibitem{Schlottmann}
P. Schlottmann, P.D. Sacramento, 
Advances in Physics, {\bf 42}, 641 (1993).

\bibitem{Affleck1993ld}
I. Affleck and A.W.W. Ludwig, 
Phys. Rev. B {\bf 48}, 7297 (1993).

\bibitem{Ludwig94}
A. W.W. Ludwig and I. Affleck,
Nucl. Phys. B {\bf 428}, 545 (1994).

\bibitem{Kimcox}
T.-S. Kim, D. L. Cox,
Phys. Rev. Lett. {\bf 75}, 1622  (1995).

\bibitem{callan90} 
C G. Callan, Jr. and L. Thorlacius, 
Nucl. Phys. B {\bf 329}, 117 (1990).

\bibitem{callan91}
C. G. Callan, Jr. and D. Freed,
Nucl. Phys. B {\bf 374}, 543 (1992).
 [hep-th/9110046]. 

\bibitem{Callan1994} C. G. Callan, 
Phys. Rev. Lett. {\bf 72}, 1968 (1994). 

\bibitem{Lee:2005ge}
T.~Lee and G.~W.~Semenoff,
JHEP {\bf 0505}, 072 (2005).

\bibitem{Hassel} 
M. Hasselfield, T. Lee, G. W. Semenoff, P. C. E. Stamp,
Ann. Phys. {\bf 321}, 2849 (2006).

\bibitem{Lee:06}
 T.~Lee,
JHEP {\bf 0611}, 056 (2006).

\bibitem{Lee2008}
T. Lee,
JHEP {\bf 02}, 090 (2008).

\bibitem{Lee2009}
T. Lee,
Int. J. Mod. Phys. {\bf A24}, 6141 (2009).

\bibitem{Lee2009q}
T. Lee, 
JHEP {\bf 03} 078 (2009).

\bibitem{LeeU(1)}
T. Lee,
{\it $U(1)$ Chiral Symmetry in One-Dimensional Interacting Electron System with Spin},
[arXiv:1510.01054] (2015).

\bibitem{LeeTL2015}
T. Lee, 
{\it The Tomonaga-Luttinger Liquid with Quantum Impurity Revisited},
[arXiv:1512.08842] (2015).
 
\bibitem{fendley1995a} P. Fendley, A. W. W. Ludwig and H. Saleur,
Phys. Rev. Lett. {\bf 74} 3005 (1995).

\bibitem{fendley1995} P. Fendley, A. W. W. Ludwig and H. Saleur,
Phys. Rev. B {\bf 52}, 8934 (1995). 

\bibitem{wong} E. Wong and I. Affleck,
Nucl. Phys. B {\bf 417}, 403 (1994). 

\bibitem{Affleck1994}
A. Affleck and W. W, Ludwig, J. Phys. A {\bf 27}, 5375 (1994).

\bibitem{chamon}
C.~Chamon, M.~Oshikawa, and I.~Affleck,
Phys. Rev. Lett.  {\bf 91}, 206403 (2003).

\bibitem{oshikawa2006}
M. Oshikawa, C. Chamon and I. Affleck,
J. Stat. Mech. P02008, 102 (2006).

\bibitem{Giuliano08}
D. Giuliano and P. Sodano,
New Jour. Phys. {\bf 10}, 093023 (2008). 

\bibitem{aristov2011}
N. Aristov and P. W\"olfle, 
Phys. Rev. B {\bf 84}, 155426 (2011).

\bibitem{aristov2013}
D. N. Aristov and P. W\"olfle, 
Phys. Rev. B {\bf 88}, 075131 (2013).

\bibitem{Shi2016}
Z. Shi and I. Affleck.
{\it A fermionic approach to tunneling through junctions of multiple quantum wires},
[arXiv:1601.00510] (2016).





\end{thebibliography}
\end{document}